\documentclass[12pt,preprint]{aastex}

\shortauthors{Kraemer et al.}
\shorttitle{Effects of Dissipative Turbulence}

\begin{document}

\title{On the Effects of Dissipative Turbulence on the Narrow Emission-Line
Ratios in Seyfert Galaxies}

\author{S.B. Kraemer\altaffilmark{1,2},
M.C. Bottorff\altaffilmark{3}, 
\& D.M. Crenshaw\altaffilmark{4}}

\altaffiltext{1}{Institute for Astrophysics and Computational Sciences,
Department of Physics, The Catholic University of America, Washington, DC
20064; kraemer@yancey.gsfc.nasa.gov}

\altaffiltext{2}{Astrophysics Science Division, Code 667, NASA's
Goddard Space Flight Center, Greenbelt, MD 20771}

\altaffiltext{3}{Physics Dept., Southwestern University, FJS123, 1001 E.
University Ave., Georgetown, TX 78626}

\altaffiltext{4}{Department of Physics and Astronomy, Georgia State University,
Atlanta, GA 30303}

\begin{abstract}

We present a photoionization model study of the effects of micro-turbulence and dissipative heating on 
emission lines for number and column densities, elemental abundances, and ionizations 
typical for the narrow emission line regions (NLRs) of Seyfert galaxies. Earlier 
studies of NLR spectra generally found good agreement between the observations
and the model predictions for most strong emission lines, such as [O~III] $\lambda$5007,
[O~II] $\lambda$3727, [N~II] $\lambda$6583, [Ne~III] $\lambda$3869, and the H and He
recombination lines. Nevertheless, the strengths of lines from species with ionization
potentials greater than that of He$^{+}$(54.4 eV), e.g. N$^{+4}$ and Ne$^{+4}$,
were often under-predicted. Among the explanations suggested for these discrepancies
were (selectively) enhanced elemental abundances and contributions from shock heated
gas. Interestingly, the NLR lines have widths of several 100 km s$^{-1}$, well 
in excess of the thermal broadening. If this is due to
micro-turbulence, and the turbulence dissipates within the emission-line gas, the gas 
can be heated in excess of that due to photoionization. We show that the combined 
effects of turbulence and dissipative heating can strongly enhance N~V $\lambda$1240 (relative to He~II $\lambda$1640), 
while the heating alone can boost the strength of [Ne~V] $\lambda$3426. We suggest
that this effect is present in the NLR, particularly within $\sim$ 100 pc of the central engine. Finally, since
micro-turbulence would make clouds robust against instabilities generated during
acceleration, it is not likely to be a coincidence that the radially outflowing emission-line
gas is turbulent.

\end{abstract}


\section{Introduction}

Seyfert galaxies are relatively nearby ($z<0.1$), low-luminosity (bolometric luminosities $L_{\rm bol} $ $<$ 10$^{45}$ erg s$^{-1}$) 
examples of Active Galactic Nuclei. Based on the widths of their optical emission lines, Seyfert
galaxies are generally divided into two types (Khachikian \& Weedman 1971). Seyfert 1 galaxies possess
broad permitted lines, with full width at half-maxima (FWHMs) $\geq$ 10$^{3}$ km s$^{-1}$, narrower
forbidden lines, with FWHM $\sim$ 500 km s$^{-1}$, and optical continua which are
dominated by non-stellar emission (e.g., Oke \& Sargent 1968).  Seyfert 2 galaxies show only 
narrow emission lines and the non-stellar contribution to their optical continua is much weaker
(Koski 1978). The broad emission lines in Seyfert 1s vary on short timescales in response to
changes in the continuum flux, which indicates that this
emission arises in dense gas within tens of light-days of the central source
(e.g. Peterson et al. 2004), in
what is referred to as the Broad Line Region (BLR). The narrow lines detected in both Seyfert 1s and 2s can
extend $\sim$ 1 kpc from the AGN (e.g. Schmitt \& Kinney 1996; Schmitt et al. 2003), forming the
so-called Narrow-Line Region (NLR). Spectropolarimetry studies revealed the presence of strongly polarized continua and
broad permitted line emission in several Seyfert 2 galaxies. This discovery led to the unified
model for Seyfert galaxies (Antonucci 1993), which posits that the differences between the two types is 
the result of viewing angle, with Seyfert 2 galaxies characterized by the obscuration
of their broad-lines regions and central engines by a dusty circumnuclear torus. In this
model, the torus collimates the ionizing radiation and the resulting illumination pattern produces a roughly
biconical NLR (Schmitt \& Kinney 1996). If the AGN is viewed at high inclination, the bicone axis is roughly in the plane of the sky,
maximizing the projected angular size of the NLR. 

While historically there has been some debate about the importance of collisional and/or shock ionization of the NLR 
(e.g., Kriss et al. 1992; Wilson \& Raymond 1999), the narrowness of Radiative Recombination Continua detected in
XMM-{\it Newton} spectra (Sako et al. 2000; Kinkhabwala et al. 2002; Turner et
al. 2003;
Armentrout, Kraemer, \& Turner 2007) is a clear indication that the emission-line gas is photoionized by
the central source. We have analyzed {\it Hubble Space Telescope (HST)}/Faint Object Spectrograph (FOS)
observations of the NLR in the Seyfert 1 galaxies NGC 5548 (Kraemer et al. 1998a) and NGC 4395
(Kraemer et al. 1999), and the Seyfert 2 galaxy NGC 1068 (Kraemer, Crenshaw \& Ruiz 1998b) and {\it HST}/Space Telescope
Imaging Spectrograph (STIS) observations of NGC 1068 (Kraemer \& Crenshaw 2000a, 2000b), the Seyfert 1 galaxy 
NGC 4151 (Nelson et al. 2000; Kraemer et al. 2000), and the Seyfert 2 galaxy Mrk 3 (Collins et al.
2005). The STIS observations were obtained in the low-resolution/long slit mode (see Woodgate
et al. 1998), with which we were able to spatially resolve the NLRs and explore the nature
of the emission-line clouds as a function of distance from the central sources. Using
photoionization models, we were able to constrain the densities, column densities, dust/gas
ratios, and ionization state of the emission-line gas. In general, we were able to fit nearly all of the 
observed emission line fluxes and ratios by assuming
the gas was photoionized solely by the central source. Nevertheless, we found discrepancies
in the predicted strengths of a few emission lines, typically those from the higher ionization
states (i.e., those with ionization potentials above the He~II Lyman limit), such as N~V $\lambda$1240\footnote{For
both the observations and the models, we consider the combined N~V $\lambda\lambda$ 1238.8, 1242.8
(N~V $\lambda$1240) and
C~IV $\lambda\lambda$1548.2, 1550.8 (C~IV $\lambda$1550), since the doublets were not resolved in the
FOS and low-resolution STIS spectra.},
[Ne~V] $\lambda\lambda$3346,3426, and [Fe~VII] $\lambda$6087. Usually, our models under-predicted the
strengths of these lines by factors of $\sim$2, although the discrepancies were often much greater
for N~V. The poor fit to high-ionization
lines was also noted by Oliva (1997), and may be rectified by adjusting the elemental abundances or including
additional ionization/excitation mechanisms (e.g. Kriss et al. 1992). In the modeling of the NLR emission
in NGC 5548 (Kraemer et al. 1998a) and NGC 1068 (Kraemer et al. 1998b), we were able to improve the fit for the N~V and
[Ne~V] lines by assuming super-solar abundances of nitrogen and neon. However, including higher abundances
for selected elements resulted in over-predictions of lines from lower ionization
states, e.g. N~IV] $\lambda$1486. Indeed, in most cases, the strengths of the lower ionization lines from these
elements, e.g., [Ne~III] $\lambda$3869 and [N~II] $\lambda\lambda$ 6548,6583, are consistent with roughly
solar abundances (see Section 2.1).

In addition to photoionization model studies of the NLR, we have also used spatially resolved STIS spectra to
analyze the kinematics of the emission-line gas in NGC 1068 (Crenshaw \& Kraemer 2000; Das et al. 2006), Mrk 3
(Ruiz et al. 2001, 2005) and NGC 4151 (Crenshaw et al. 2000; Das et al. 2005). Although the kinematic studies utilized 
the [O~III] $\lambda$5007 line, we have found that the higher ionization lines are spatially co-located
with the [O~III] knots
(e.g. Collins et al. 2005), therefore it is probable that similar dynamical effects drive the higher
ionization gas. Each of these sources
show a similar kinematic profile, in which the projected radial velocity, $v_{r}$, gradually increases from the central point
source out to $r$ $\sim$ 100 pc, after which the velocities abruptly begin to decrease towards systemic. We have been able to 
model the kinematics by assuming the gas is confined to a hollow, bi-conical envelope, whose apex is roughly coincident
with the AGN. The gas is continuously accelerated, then begins to undergo a rapid deceleration. The dynamics of this process
is still not clear (e.g. Everett \& Murray 2006; Das, Crenshaw, \& Kraemer 2007), but it is likely that radiation pressure from the continuum source 
makes an important contribution. The [O~III] lines are also quite broad in these objects, e.g., FWHM $>$ 1000 km s$^{-1}$ near the
apex of the bicone and $\gtrsim$ 500 km s$^{-1}$ throughout the inner 200 pc of the NLR. In general, the widths appear
to decrease with distance from the AGN, although there are points of abrupt increases, particularly in the case of
NGC 1068 (Crenshaw \& Kraemer 2000; Das et al. 2006), which occur near the velocity turn-over points. The physical conditions
that produce these line widths is unclear (see Section 5.). It is possible that they result from the super-position of multiple
individual kinematic components; if so, the flow may simply become more chaotic at the turn-over points, broadening
the profiles. However, it is also possible that the emission-line knots possess significant micro-turbulence. Whatever
process slows the knots could conceivably boost the micro-turbulence. Another piece of evidence that the outflowing gas
is turbulent is found in the blue-shifted UV absorption detected along the line-of-sight to many Seyfert 1 galaxies
(Crenshaw et al. 1999; Crenshaw, Kraemer, \& George 2003; Dunn et al. 2007). The absorption lines nearly always
have widths ($>$ tens of km s$^{-1}$) in excess of 
thermal ($\sim$ several km s$^{-1}$) and, in several cases, we have measured FWHMs  $>$ several hundred km s$^{-1}$ (e.g. Kraemer et al. 2001). 
The absorption lines appear/disappear
over short timescales (e.g. Crenshaw \& Kraemer 1999; Kraemer, Crenshaw, \& Gabel 2001; Crenshaw et al. 2003) without obvious changes in
their widths, which is suggestive of turbulent knots passing across our line-of-sight.

In Figures 1 -- 3, we show the N~V $\lambda$1240/He~II $\lambda$1640,  C~IV $\lambda$1550/He~II $\lambda$1640, and
[Ne~V] $\lambda$3426/He~II $\lambda$1640
ratios as a function of projected radial distance, $r$, and FWHM. 
There is some evidence for higher values at $r \lesssim 100$ pc, although there are some points with
high ionization ratios at greater
radial distances for C~IV and [Ne~V]. More evident is the correlation of the line ratios with FWHM. 
If the broader lines are found in more highly excited components, as the ratios indicate, it suggests
some physical connection. Therefore, our hypothesis is that the line widths result, at least partly, from internal micro-turbulence, which
in turn affects the observed line ratios.

A related issue is that the broad emission lines in AGN are smooth (e.g., Dietrich, Wagner, \&
Courvoisier 1999). If the lines are
thermally broadened, there must an inordinately large number of individual clouds within the BLR.
To address this problem, Bottorff \& Ferland (2000) proposed that the clouds are internally turbulent, with
micro-turbulent velocities that exceed their thermal velocities. Using photoionization
models, they were able to generate
smooth line profiles from a small number of clouds. Bottorff \& Ferland (2002, hereafter BF2002) expanded on this
by examining the effects of dissipative turbulence. In the case of non-dissipative
turbulence, the turbulent motions persist, whereas, in the dissipative case, the motion is converted
into heat. The corresponding increase in electron temperature will affect the emissivity of emission
lines, particularly those that are collisionally excited. There is evidence for dissipative
heating in the diffuse interstellar medium of the Milky Way Galaxy (Minter \& Balser 1997), hence
BF2002 argued that such processes were likely to occur within the BLR gas
in AGN. Since the thermal motions will decay due to the heating losses, there must be some
continual external source of turbulence. Possibilities include the intense radiation pressure
experienced by gas close to the AGN and magnetic fields. Interestingly, both radiation pressure
and magneto-hydrodynamic (MHD) Flows
have been suggested as mechanisms for mass loss in AGN (see Crenshaw et al. 2003, and
references therein), hence it is possible that forces that drive the outflows also maintain
the micro-turbulence. 

BF2002 parameterized the dissipative heating as a function of turbulence velocity, as
follows:
\begin{equation}
Q = \eta_{v} \rho \frac{v_{t}^{3}}{D} {\rm ergs}~ {\rm cm}^{-3}~ {\rm s}^{-1},
\end{equation}
where $\eta_{v}$ is of order unity (see Stone, Ostriker, \& Gammie 1998), $v_{t}$ is the magnitude of the turbulence velocity\footnote{The turbulence velocity is equivalent to the velocity dispersion, $\sigma (v)$, which
is 1/2.355 times the Full Width Half Maximum (FWHM) of a Gaussian line profile.}, 
$\rho$ is the mass density of the gas, and $D$ is the scale length over which the turbulence dissipates.
BF2002 assumed that $D$ corresponds
to the physical depth of a BLR cloud, or $\sim$ 10$^{13}$ cm, based on typical BLR hydrogen number densities
($n_{H}$ $\sim$ 10$^{10}$ cm$^{-3}$; Davidson \& Netzer 1979) and column densities ($N_{H}$ $\sim$ 10$^{23}$
cm$^{-2}$; Kwan \& Krolik 1981). For their models, they re-parameterized the heating rate as:
the following:
\begin{equation}
Q \approx 2.3 \times 10^{-3} v_{3}^{3} \frac{n_{10}}{L_{13}} {\rm ergs}~ {\rm cm}^{-3}~ {\rm s}^{-1},
\end{equation}
where $v_{3}$ is the turbulence velocity in units of 1000 km s$^{-1}$, $n_{10}$ is the hydrogen
number density in units
of 10$^{10}$ cm$^{-3}$, $L_{13}$ is the physical depth of the cloud, in units of
10$^{13}$ cm. 
In this paper, we examine the effects of dissipative turbulence in the NLR gas and
have used equation (2) to calculate the heating.  Following BF2002, we assumed that the turbulence dissipates
over the full depth of the cloud. However, although the column densities of NLR tends to be
much lower than the canonical BLR value (Kraemer et al. 1998a, 1998b; Kraemer et al. 2000, Kraemer \& Crenshaw 2000a, 2000b),
the physical depths are several orders of magnitude larger, due to the much lower densities.

\section{Photoionization Modeling}

We have calculated a set of photoionization models to investigate the effects
of micro-turbulence and dissipative heating on the NLR emission-line ratios. The
elemental abudances and other model input parameters are detailed in the next two
subsections.

\subsection{Elemental Abundances}

The relative strengths of emission lines strongly depend on the elemental abundances. Since 
the thermal equilibrium in the optical emission-line gas is driven largely by collisional cooling,
the electron temperature is quite sensitive to the relative fraction of heavy atoms (neutral and ionized) in gas phase. For example,
for heavy-element abundances more than several times solar values, the collisionally excited lines
can actually be weaker than for lower abundances due to the increased line-cooling rates.
While the abundances of common elements such as C, O, and Ne scale with $Z/Z_\odot$ (where $Z_\odot$
indicates  ``solar abundances''), N scales as ($Z/Z_\odot$)$^{2}$ (e.g., Vila-Costas \& Edmunds 1993). Hence, N/C-O-Ne can become quite large
at super-solar abundances, which should result in the enhancement of lines such as N~V $\lambda$1240 and 
the N~IV] $\lambda$1486 multiplet compared to lines from these other elements.

In our previous NLR studies (see Section 1), we have assumed ``roughly solar'' abundances (e.g. Grevesse \& Anders 1989). Recently, 
Groves, Dopita, \& Sutherland
(2004) re-visited the issue of elemental abundances in the NLR, and incorporated the latest efforts
to define solar abundances (e.g. Grevesse \& Sauval 1998), which suggested somewhat lower
N/H than previous studies. In order to reproduce the observed strengths of the [N~II] $\lambda\lambda$ 6548,6583 lines
with their photoionization models, Groves et al. had to assume twice the solar N/H ratio (interestingly, this
is quite close to ``roughly'' solar value that we had used in our previous NLR modeling).
It is worth noting that similarly elevated abundances were determined for the blue-shifted UV absorbers
in the Seyfert 1 galaxy Mrk 279 (Arav et al. 2007), particularly since the UV absorbers and 
the emission-line gas in the inner NLR may be associated (Crenshaw \& Kraemer 2005). Assuming all the
elements have scaled together, this corresponds to a overall abundance of  $Z/Z_\odot$ $\approx$ 1.4.
For this paper, we used the abundances from Asplund, Grevesse, \& Sauval (2005), scaled in
manner suggested by Groves et al. The logs of the abundances relative to H by number are: He: $-1.0$;
C: $-3.46$: N: $-3.92$; O: $-3.19$; Ne: $-4.01$; Mg: $-4.33$; Si: $-4.34$; S: $-4.69$; Fe: $-4.5$. 
Note that the Ne/O ratio is $\sim$ 0.15, in agreement with current estimates that include the
effect of solar CNNe mixing (see Delahaye \& Pinsonneault 2006).     

Although much of the NLR gas may be dusty (Kraemer \& Harrington 1986; Groves et al. 2004) we have not included dust in 
these models. The reasoning is twofold. First, there would be no depletion of neon and little if any of nitrogen (in the
absence of ice mantles) onto grains. Second, we have found evidence that the dust/gas ratios in the NLR are below
that of the ISM and, in some cases (e.g. NGC 1068: Kraemer \& Crenhsaw 2000a), the gas in the inner NLR appears to be
dust-free.

\subsection{Model Input Parameters}

The photoionization models used for this study were generated using the Beta 5
version of Cloudy (Ferland et al. 1998).
As per convention, the models are parameterized in
terms of the ionization parameter, $U = Q/(4\pi r^{2} c n_{H})$
where $r$ is the distance between the emission-line gas and the central source,
$n_{H}$ is the hydrogen number density, $c$ is the speed of light, and $Q = \int_{13.6 eV}^{\infty}(L_{\nu}/h\nu)~d\nu$,
or the number of ionizing photons s$^{-1}$ emitted by a source of luminosity
$L_{\nu}$ ergs s$^{-1}$ Hz$^{-1}$. We assumed a plane-parallel (``slab'') geometry.
For the incident continuum, we used the spectral energy distribution which we employed
in our modeling of NGC 4151 (Kraemer et al. 2005), which is parameterized as a broken power law of the form $L_{\nu} \propto 
\nu^{\alpha}$ as follows: $\alpha = -1.0$ for energies $<$ 13.6 eV,
$\alpha = -1.45$ over the range 13.6 eV $\leq$ h$\nu$
$<$ 0.5 keV, and $\alpha = -0.5$ above 0.5  
keV. We included a low energy cut-off at $1.24 \times 10^{-3}$ eV (1 mm) and a high energy cutoff
at 100 keV. While we compare our model predictions with emission-line spectra from other Seyfert galaxies,
the results were not sensitive to small changes in the power-law indices or break energies, and
using a single model SED makes comparison of the model predictions to the complete set of
emission-line ratios more straightforward.

Based on our previous models, NLR gas within a few tens of pcs of the AGN can be
characterized assuming $n_{H}$ $=$ 10$^{5}$ cm$^{-3}$. 
The line ratios of interest to this study do not change appreciably for lower values of $n_{H}$, and the
value used is well below the critical density for the $^{1}D_{2}$ level of Ne V (the upper level
of the 3426 \AA~transition; Osterbrock 1989). We assumed $N_{H}$ $=$ 10$^{21}$ cm$^{-2}$, which
is consistent with matter-bounded clouds over the range of ionization we used, i.e.,
$-2 \leq {\rm log}(U) \leq 0$.  Based on $n_{H}$ and $N_{H}$, the physical depth for the model clouds was 10$^{16}$ cm. The 
turbulence velocity, $v_{t}$, is
directly command-able in Cloudy (Ferland et al. 1998). The dissipative heating, based on the
density, physical depth and $v_{t}$ (see Table 1), was introduced 
via the command-able additional heating term. We examined the effects of turbulence and
dissipative heating for a range of  $v_{t}$ from 0 (thermal broadening only) to 250 km s$^{-1}$. 

\section{Model Results}

In our models, we tested the effects of turbulence/heating on two emission lines that are strong and often
under-predicted by photoionization models: N~V $\lambda$1240 and [Ne~V] $\lambda$3426.
As shown in Figure 4, the He$^{+2}$ zone, in which there is the strongest contribution to the
He~II recombination lines, 
is coincident with the N$^{+4}$ and Ne$^{+4}$ zones. Therefore, the results are given as line
ratios relative to the He~II $\lambda$1640 recombination line.

For resonance lines such as N~V $\lambda$1240, there can be a contribution from photo-excitation to the
upper level by continuum radiation. This has been suggested as the mechanism responsible
for the strong resonance lines from He-like and H-like ions of Ne, O, and N detected in X-ray spectra 
of Seyfert galaxies (e.g. Sako et al. 2000; Kinkhabwala et al. 2002; Armentrout et al.
2007). If the gas is turbulent, the absorption profiles are broadened, which increases the number of
continuum photons that the gas can absorb before becoming optically thick to the incident radiation,
thereby increasing the contribution from photo-excitation. To test the effect of photo-excitation alone, 
we generated a set of models with turbulence but no dissipative heating. The results for N~V are
shown in Figure 5. In the absence of turbulence, the maximum N~V/He~II ratio is $\sim$ 1.4, which
occurs near an ionization parameter of log$(U) = -0.4$.  The N~V/He~II ratio increases with turbulence, reaching
a value of $\sim$ 4 for $v_{t}$ $=$ 250 km s$^{-1}$. The shape of the turn-over at log$(U) > -0.4$  
depends
on the predicted He~II $\lambda$1640, which is relatively stronger at high turbulence as the
model diverges from the Case B approximation (e.g. Osterbrock 1989). Therefore, elevated N~V/He~II
ratios can be achieved for large micro-turbulence, although there is no effect on
forbidden lines such as [Ne~V] $\lambda$3426. However, the range in ionization parameter for which
N~V/He~II ratios $>$ 2 are predicted is rather narrow and covers a range for which the gas is
so highly ionized that most of the nitrogen is in higher ionization states than N$^{+4}$. The N~V is enhanced
in this case because the smaller N$^{+4}$ column densities result in lower N~V optical depths. 
The effect of turbulence alone is not as strong at the ionization parameter at which
the largest column density of N$^{+4}$ is predicted, log$(U) \approx - 1.3$, since the line becomes optically thick close to the
irradiated face of the slab. Also, the NLR
must include gas at lower
ionization, which still produces He~II emission, hence the contribution from turbulent gas with high N~V/He~II ratios will 
be diluted. Therefore, photo-excitation, even when maximized by high turbulence, may not be sufficient to produce
the highest observed NV~/He~II ratios (see Figure 1), unless the nitrogen abundances are $\gtrsim$ several times solar.

The effect of dissipative heating scaled with micro-turbulence on the N~V/He~II ratio is shown in
Figure 6. The peak ratio increases with heating, peaking at $\sim$ 11 for 
$v_{t}$ $=$ 250 km s$^{-1}$. Comparing these results to the models generated without heating,
in addition to predicting high ratios over a broader range of ionization parameter, the ionization
parameters at which N~V/He~II peaks drop with increased heating/turbulence. This demonstrates the
relative contributions from photo-excitation and dissipative heating: when 
heating is included, there is a significant enhancement of the line in the more optically thick models, and
the heating becomes the dominant effect for $v_{t}$ $>$ 200 km s$^{-1}$.
Including heating has a significant effect on the [Ne~V]/He~II ratio, as well, as shown in Figure 7. The slight trend to
lower values of $U$ for the peak ratio is due to the higher electron temperatures for the models with the
greatest Ne$^{+4}$ fraction. As these results indicate, [Ne~V]/He~II ratios greater than unity require
the heating resulting from $v_{t}$ $>$ 100 km s$^{-1}$.  

Since we have included dissipative heating, we examined the thermal stability of the gas. The thermal
stability curves (e.g., Krolik, McKee, \& Tarter 1981) are shown in Figure 8. Although the unstable region, characterized by negative slope, is 
more pronounced for the models with 
$v_{t}$ $=$ 250 km s$^{-1}$, the gas will be thermally stable for the range of 
ionization parameters from which we examined emission-line ratios, i.e., log$(U) < 0.0$.

\section{Comparison with Observations}

Since the under-prediction of the
N~V/He~II and [Ne~V]/He~II ratios is the motivation for this study, we now demonstrate to what
degree the addition of dissipative heating can reduce these discrepancies. As discussed
in Section 1, kinematic studies have found that the [O~III] $\lambda$5007 lines in the NLR can
be broad. For example, the brightest components in NGC 1068 typically have FWHM of
$\lesssim$ 700 km s$^{-1}$ (Das et al. 2006), while those in NGC 4151 have FWHM of 
$\lesssim$ 500 km s$^{-1}$ (Das et al. 2005). Also, the largest FWHM detected among the strong 
UV absorption components in NGC 4151 is $\approx$ 435 km s$^{-1}$ (Kraemer et al. 2001). 
In order to test the effects of micro-turbulence, we have opted to be somewhat
conservative and have generated models with turbulence velocities of, 50, 100, and 150 km s$^{-1}$ (which
corresponds to FWHM $\approx$  120, 235, and 350 km s$^{-1}$) for comparison to the observed line ratios.
The values assumed for $U$, $N_{H}$, and $n_{H}$ are as described in Section 2.2.  
      
In Figure 9, we compare the model predictions for N~V $\lambda$1240/He~II $\lambda$1640 versus
[Ne~V] $\lambda$3426/He~II $\lambda$1640 against the observations. Although there are a large number of datapoints for which
both the N~V and [Ne~V] lines are relatively weak, it is apparent that turbulence combined with dissipative heating is required
for those points with N~V/He~II $>$ 1.5 and [Ne~V]/He~II $>$ 0.5. The few points that lie outside the
upper ($v_{t}$ $=$ 150 km s$^{-1}$) curve could be matched by higher turbulence/heating. In particular, note the 
datapoint for NGC 5548. In Kraemer et al. (1998a), we had suggested that the large N~V/He~II and [Ne~V]/He~II ratios
in the source might be
the result of a high nitrogen and neon abundances. However, these results  
suggest that strong N~V and [Ne~V] lines are more likely indicative of
turbulence/heating.

In Figure 10, we show the model predictions for N~V/He~II versus C~IV $\lambda$1550/He~II $\lambda$1640.
The model
comparison shows that C~IV/He~II ratios $>$ 4 are consistent with turbulence/heating. We also compared
the N~V/He~II is to N~IV] $\lambda$1486/ He~II. While the number of datapoints was small due to the relative
weakness of N~IV], combined with the effects of reddening, there is some indication that the N~IV] may show the effects
of dissipative heating.

An additional indication that the NLR gas is turbulent is that high L$\beta$/L$\alpha$ ratios
have been detected in some Seyferts. 
In {\it Hopkins Ultraviolet Telescope} spectra of NGC 1068, Kriss et al. (1992) measured a ratio of $\sim$ 0.1,
while, under Case B
conditions (Osterbrock 1989), this ratio will be $<< 0.01$. In our models, we find that 
for log$(U)$ $=$ $1.5$, without turbulence, L$\beta$/L$\alpha$ $\approx$ 1.4
$\times$ 10$^{-3}$. 
For $v_{t}$ $=$ 150 km s$^{-1}$, the ratio is $\approx$ 0.01. Decreasing the column density
by a factor of 10 increases L$\beta$/L$\alpha$ by a similar factor. Hence, this strongly
suggests that the gas in the  NLR
is turbulent, although the column densities of the Lyman-emitting gas may be somewhat lower
than those assumed here, but well-within the range of values used in our
previous NLR studies (e.g. Kraemer et al. 2000; Kraemer \& Crenshaw 2000b).
 
\section{Discussion}

 The inclusion of micro-turbulence with associated dissipative heating has the 
effect of increasing the strengths of the NV and [NeV] lines, relative to HeII. 
In the range of ionization parameter for which these lines are strongest, i.e. 
$-1.5 < {\rm log}(U) < -1.3$, the volume heating rate for a turbulence velocity $=$ 150 
km s$^{-1}$ is on the same order as the radiative heating rate.
Spatially resolved spectra of the NLR 
of Seyferts (see Section 1) reveal that the 
emission-line knots remain quite broad (FWHM $>$ several
hundred km s$^{-1}$) throughout the inner $\sim$ 200 pc of the NLR.
This suggests that if this is due to micro-turbulence, the turbulence is being continuously driven, perhaps
by the continuum radiation from the central source. For example, in NGC 4151,
which has a luminosity in ionizing photons of $\sim$ 10$^{53}$ photons s$^{-1}$
(Kraemer et al. 2005), an emission-line cloud, characterized by log$(U) = -1.5$,
$n_{H} = 10^{5}$ cm$^{-3}$, and $N_{H}$ $=$ 10$^{21}$ cm$^{-2}$, would lie at $r = 9.2 \times 10^{18}$ cm. If this gas formed
a complete shell around the central source, to produce the dissipative heating rate
assumed in our models (see Table 1) would require an energy input rate of 8.2 $\times$ 10$^{41}$
erg s$^{-1}$, which is a fraction of the bolometric luminosity of
NGC 4151, even during the low-flux state observed in 2002 May ($L_{bol}
\approx 5 \times 10^{43}$ erg s$^{-1}$; Kraemer et al. 2005). Therefore, {\it based on
the energetics}, it is possible that the continuum radiation can continuously drive the 
micro-turbulence.

Since the emission-line gas is part of a radial outflow, one must consider possible
connections between the line widths and the means by which the gas is accelerated. 
Various mechanisms have been suggested for 
the acceleration including radiation pressure 
(Blumenthal \& Mathews 1975; Mathews 1976), ram pressure from winds (Weymann et 
al. 1982; Schiano 1986; Smith 1993) and MHD-driven flows (Blandford \& Payne 1982; 
Bottorff, Korista, \& Shlosman 2000). A Blandford \& Payne type MHD wind has
(in cylindrical coordinates, $r$, $z$, $\phi$) a $z$-component of
outflow velocity that increases monotonically, 
but the $r$- and $\phi$- velocity components can rapidly increase (e.g., in the BLR) 
but then decrease further out (e.g., in the NLR). Therefore it is possible 
to have large NLR line widths near the base and small line widths far from the base 
that have nothing to do with turbulence or an outwardly propagating disturbance.
Our kinematic studies have shown that the emission-line gas is accelerated to 
velocities $\sim$ 1000 km s$^{-1}$ within the inner 100 pc of the NLR. Over the same 
distance scales, the FWHM of the lines decrease from 1000 km s$^{-1}$ to a few 
hundred km s$^{-1}$ (with some notable exceptions). Although this is consistent with 
super-position of velocity components, taken together with the correlation between the
N~V/He~II, C~IV/He~II, and [Ne~V]/He~II ratios and FWHM of these lines, we   
interpret this as evidence that it is turbulence (largely) 
and not systematic non-radial components of the outflow that contribute to the
line widths.

While a cloud of gas is being accelerated, the driving 
mechanism will generate instabilities within the cloud, such as Rayleigh-Taylor, 
for a radiatively driven cloud, or Kelvin-Helmholtz instabilities, for a cloud
entrained in a wind. If the timescale for 
the instability to propagate through the cloud is less than the timescale for 
acceleration, e.g. to achieve the observed radial velocities, the cloud will 
fragment. It has been shown (e.g. Allen 1984; Smith 1993) that fragmentation is 
inevitable for clouds with physical parameters similar to those derived from our 
photoionization models (which are the basis for the range in parameters assumed 
in this paper). There are only two ways to reconcile this. One way is for the 
clouds to constantly form and evaporate within a hot wind (e.g. Krolik \& Kriss 
2001). The other way is for the clouds to be robust to instability-driven 
fragmentation via internal turbulence (Allen 1984). The turbulence could be 
generated by multiple shocks which also accelerate the clouds (Marscher 1978). 
Interestingly, this requires turbulence velocities $0.1 v_{r} < v_{t} < 
v_{r}$, which is consistent with observations; for $v_{t} > v_{r}$, the 
clouds will evaporate. Marscher (1978) further notes that stable clouds must 
have radii $>$ 10$^{14}$ cm, which is consistent with our modeling results.

Micro-turbulence in NLR clouds will dissipate into heat on a time scale 
shorter than the NLR crossing time. This means that energy must be 
provided constantly to drive the turbulence. As mentioned above it is 
energetically possible for the continuum to do this. However 
transferring the energy into turbulent motions at the required rate may 
more problematic. If the outflow is an MHD wind magnetic field lines 
which thread NLR clouds also thread the portion of the disk from which 
the NLR material was originally launched. This provides a possible 
energy conduit to the NLR clouds. It is uncertain however whether 
disturbances launched from the disk could travel over a distance of 100 
pc to be dissipated in NLR clouds.  A combination of radiative driving 
and an MHD wind however offers the possibility that  NLR materials can 
be driven against magnetic field lines thereby exciting disturbances 
which would dissipate more or less locally as heating. Such a mechanism 
is an area for future study. At the very least the presence of magnetic 
fields may be necessary to avert overly strong shock formation in the highly turbulent 
gas.

  Therefore, we suggest the following scenario. The emission-line clouds form relatively
close to the central source, perhaps in the inner few pcs of the NLR. They are accelerated 
outwards by one of the mechanisms listed 
above. In the process, they develop significant micro-turbulence, which also 
serves to maintain them against disruption. As the clouds move outwards, the 
mechanism which generates the turbulence becomes weaker, which could be due the 
$1/r^{2}$ dilution of the radiation, or a drop in density of a wind. The
dissipative heating would decrease, and there would be less enhancement of these
emission-line ratios, as is observed (see Figures 1 -- 3).

\section{Summary} 

We have explored the effects of micro-turbulence and associated dissipative
heating on emission-line ratios for conditions applicable to the NLR of
Seyfert galaxies. Based on our modeling results, we suggest the following:

1. Micro-turbulence can increase the effect of photo-excitation,
thereby boosting the relative strengths of resonance lines such as
N~V $\lambda$1240. Turbulence alone will not directly affect collisionally excited, forbidden
lines, such as [Ne~V] $\lambda$3426. However, we find that, if the turbulence dissipates over scale-lengths equal to the
physical depths of the emission-line clouds, the resulting heating is sufficient
to explain the highest [Ne~V]/He~II ratios detected in these spectra. 

2. Although we found that the elemental abundances may be slightly super-solar, 
when turbulence and heating are included, there is no need to
invoke higher abundances to explain the observed line ratios.  

3. Although we cannot rule out that the emission-line widths may partly be due
to super-position of kinematic components, the correlation between the line ratios
and FWHM suggests that the clouds are turbulent and that the turbulence affects the
relative strengths of the emission lines. This is further supported by the fact that including 
micro-turbulence of the same order of magnitude as the observed line widths can alleviate
discrepancies between the models and the observations. While the turbulence appears to diminish with radial
distance, the fact that the line widths exceed thermal throughout the 
inner $\sim$ 200 pc of the NLR suggests
that some process maintains the turbulence over large distances.

4. Micro-turbulence can make the clouds robust to
the instabilities generated during cloud acceleration. This suggests that the
lifetime of non-turbulent clouds in the inner NLR may be brief. Hence,
it may not be a coincidence that the emission and absorption lines
observed in Seyfert galaxies have widths that far exceed their
thermal widths. 

In this paper, we have not attempted to fully explore the range of parameters, i.e.,
density, ionization, elemental abundances, and dust/gas ratio, that
may exist in the NLR of Seyferts. Instead, we simply attempted to 
gauge the effects of turbulence/heating. In future work, it will be critical to examine
the origins of the turbulence and develop a more sophisticated model for
the dissipative heating, in particular regarding scale-lengths and the 
effects of internal magnetic field. In the meantime, these results suggest that it would be
appropriate to include an approximation of turbulence/heating in photo-ionization
studies of AGN. In a future paper, we plan to re-analyze the STIS long-slit spectra of NGC 1068 and
explore the role of turbulence in detail.

\acknowledgments
This
research has made use of NASA's Astrophysics Data System. We thank Gary Ferland
for his continued efforts in developing and upgrading the photoionization
code Cloudy. We thank Bruce Woodgate and Fred Bruhweiler for useful discussions.
We thank an anonymous referee for their useful comments.

\clearpage

\begin{deluxetable}{ll}
\tablecolumns{2}
\footnotesize
\tablecaption{Model Parameters$^{a}$}
\tablewidth{0pt}
\tablehead{
\colhead{Turbulence Velocity (km s$^{-1}$)} & \colhead{Heating$^{b}$ (ergs s$^{-1}$ cm$^{-3}$)}
}
\startdata
50 & 2.9 $\times$ 10$^{-15}$ \\
100 & 2.3 $\times$ 10$^{-14}$\\
150 & 7.8 $\times$ 10$^{-14}$\\
200 & 1.8 $\times$ 10$^{-13}$\\
250 & 3.6 $\times$ 10$^{-13}$\\
\enddata
\tablenotetext{a}{For the set of models, we assumed $N_{H}$ $=$ 10$^{21}$ cm$^{-2}$,
$n_{H}$ $=$ 10$^{5}$ cm$^{-3}$, and the elemental abundances listed in Section 2.1}
\tablenotetext{b}{For comparison, the radiative heating is 9.5 $\times$ 10$^{-14}$ ergs s$^{-1}$ cm$^{-3}$ and
1.1 $\times$ 10$^{-13}$ ergs s$^{-1}$ cm$^{-3}$, for log$(U) = -1.5$ and $-1.0$, respectively.}
\end{deluxetable}

\clearpage
\figcaption[nv.eps]{Dereddened NV $\lambda$1240/He~II $\lambda$1640 ratios 
measured in {\it HST}/STIS long-slit spectra of NGC 1068 (Kraemer \& Crenshaw
2000a, 2000b) as a function of radial distance projected onto the plane of the sky(top panel) and FWHM (bottom panel). The
filled and unfilled squares are from blue- and red-shifted knots, respectively.  
Note the increase in the line ratios with increasing FWHM.}

\figcaption[civ.eps]{Dereddened  C~IV $\lambda$1550/He~II
$\lambda$1640 ratios for NGC 1068.}
 
\figcaption[nev.eps]{Dereddened [Ne~V] $\lambda$3426/He~II
$\lambda$1640 ratios for NGC 1068.}

\figcaption[fracnew.ps]{Selected fractional ionic abundances as a function of column density for an ionization
parameter log$(U) = -1.0$. Note that the fractions of N$^{+4}$ and Ne$^{+4}$ become negligible for
${\rm log} N_{H}$ $>$ 21.5, at which point the gas has become optically thick above the He~II Lyman limit.}

\figcaption[nv_v_u_noQ.ps]{The ratio of N~V $\lambda$1240/He~II $\lambda$1640 as a function of $U$, for 
micro-turbulence velocities $v$ = 0 (solid line), 50 (long-dash), 100 (dash triple-dot), 150 (dash-dotted)
200 (short-dashed), and 250 (dotted) km s$^{-1}$. Dissipative heating was not included in these models,
hence the enhancement of the N~V is results solely from photo-excitation.}

\figcaption[nv_v_u2.ps]{Same as Figure 5, but dissipative heating has been incorporated in the models (see Table 1.).
Note that 1) the maximum N~V/He~II ratio predicted is a factor of nearly 3 times that for the turbulence-only models, and 2)
the peak shifts to lower values of $U$ as a function of $v$, as a result of the increased dissipative heating
at high turbulence velocities.}

\figcaption[nev_v_u.ps]{Same as Figure 6, showing the [Ne~V] $\lambda$3426/He~II $\lambda$1640 ratio. As noted in the text,
for the densities and elemental abundances used in these models, [Ne~V]/He~II ratios $>$ unity require 
the heating predicted for $v$ $>$ 100 km s$^{-1}$.}

\figcaption[scurve_new.ps]{Thermal stability curves for our set of model parameters. The three curves are: no turbulence (solid
line), $v_{t} = 150$ km $s^{-1}$ (dotted line), and $v_{t} = 250$ km $s^{-1}$ (dashed line). The asterisks correspond to
log$(U) = -1.4$, at which [Ne~V]/He~II peaks, and the crosses correspond to log$(U) = 0.0$}

\figcaption[nv_nev_new2.ps]{Predicted N~V $\lambda$1240/He~II $\lambda$1640 versus [Ne~V] $\lambda$3426/He~II $\lambda$1640
are shown for models including dissipative heating for turbulence velocities 50 (dotted), 100 (dashed), and
150 (dash-dotted) km s$^{-1}$. The curves are generated for a range of ionization parameters $-2.0 \leq {\rm log}(U) \leq 0.0$, 
running counterclockwise, with the log$(U)$ values marked at several points for the $v_{t} = 100$ and 150 km s$^{-1}$ models. The observed ratios are from our previous NLR studies; the symbols from the different datasets are 
NGC 4151, crosses (Kraemer et al. 2000), NGC 1068, asterisks (Kraemer \& Crenshaw 2000b), Mrk 3, squares
(Collins et al. 20005), NGC 5548, triangle (Kraemer et al. 1998a), and the NGC 1068 ``Hot Spot'', diamond (Kraemer \& Crenshaw
2000a).}
 
\figcaption[nv_civ_new.ps]{Predicted N~V $\lambda$1240/He~II $\lambda$1640 versus C~IV $\lambda$1550/He~II $\lambda$1640
are shown for models including dissipative heating.  Symbols are as described in Figure 9. }

\clearpage
\begin{figure}
\vspace*{-10mm}
\plotone{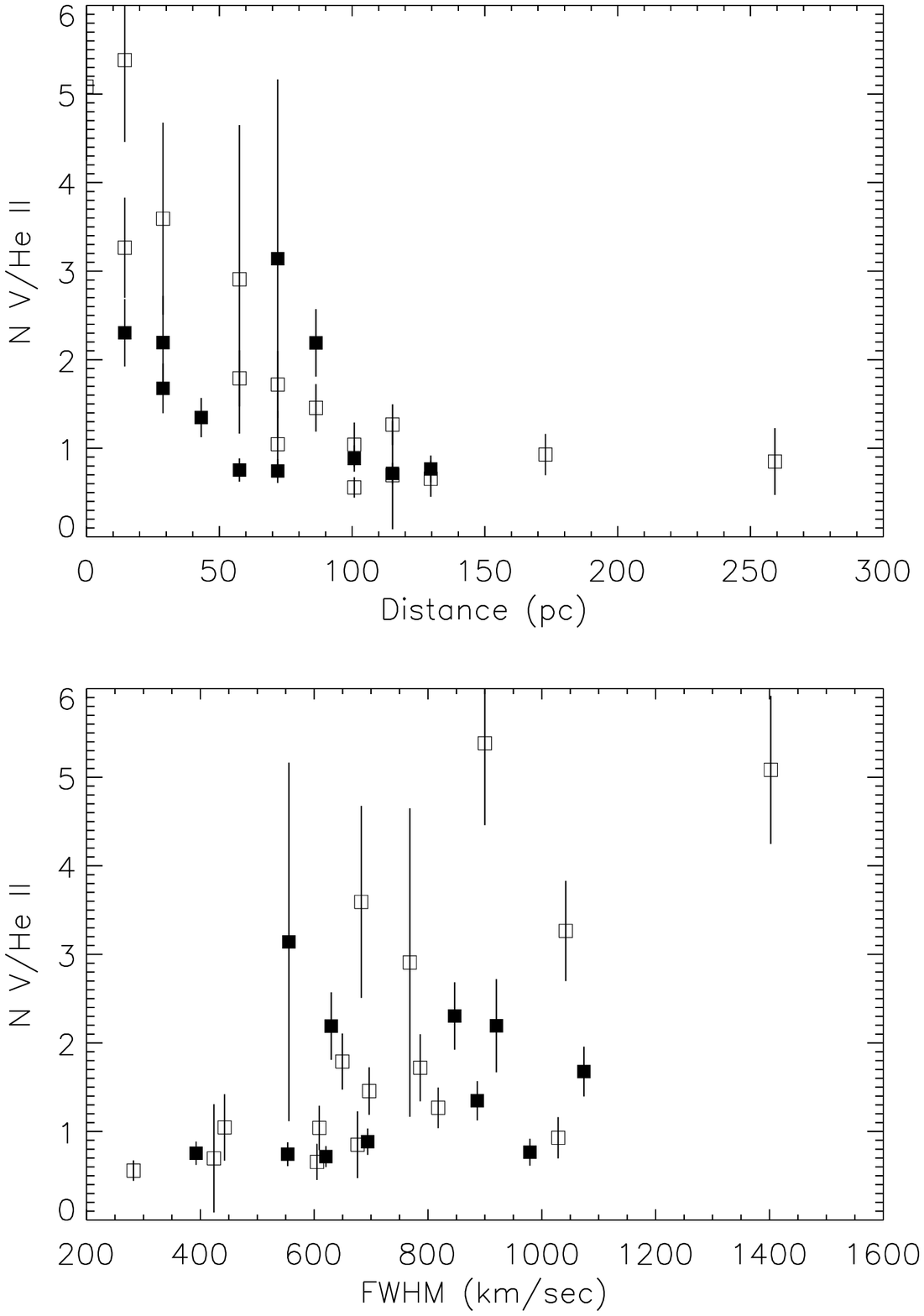}
\\Fig.~1.
\end{figure}

\clearpage
\begin{figure}
\vspace*{-10mm}
\plotone{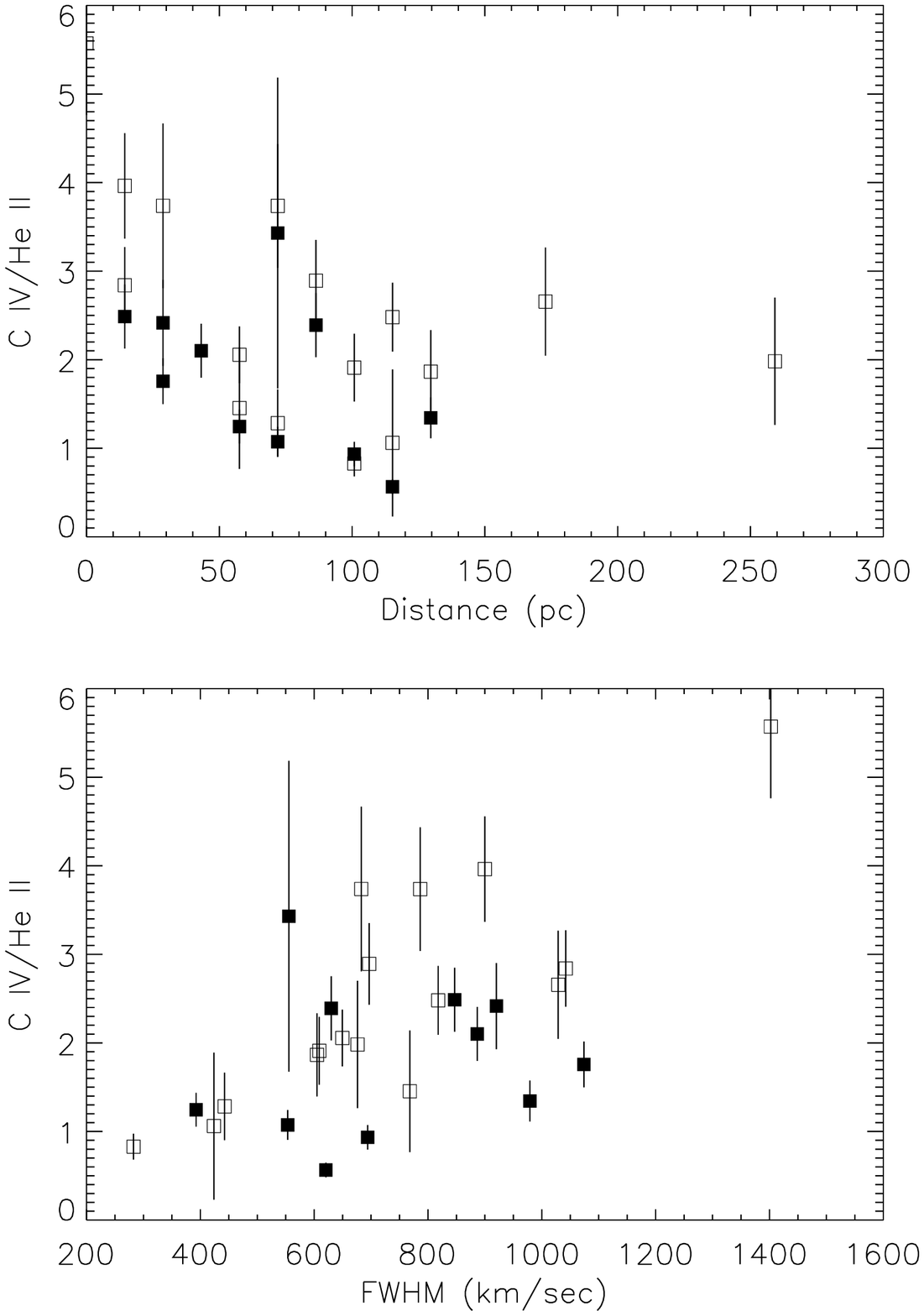}
\\Fig.~2.
\end{figure}

\clearpage
\begin{figure}
\vspace*{-10mm}
\plotone{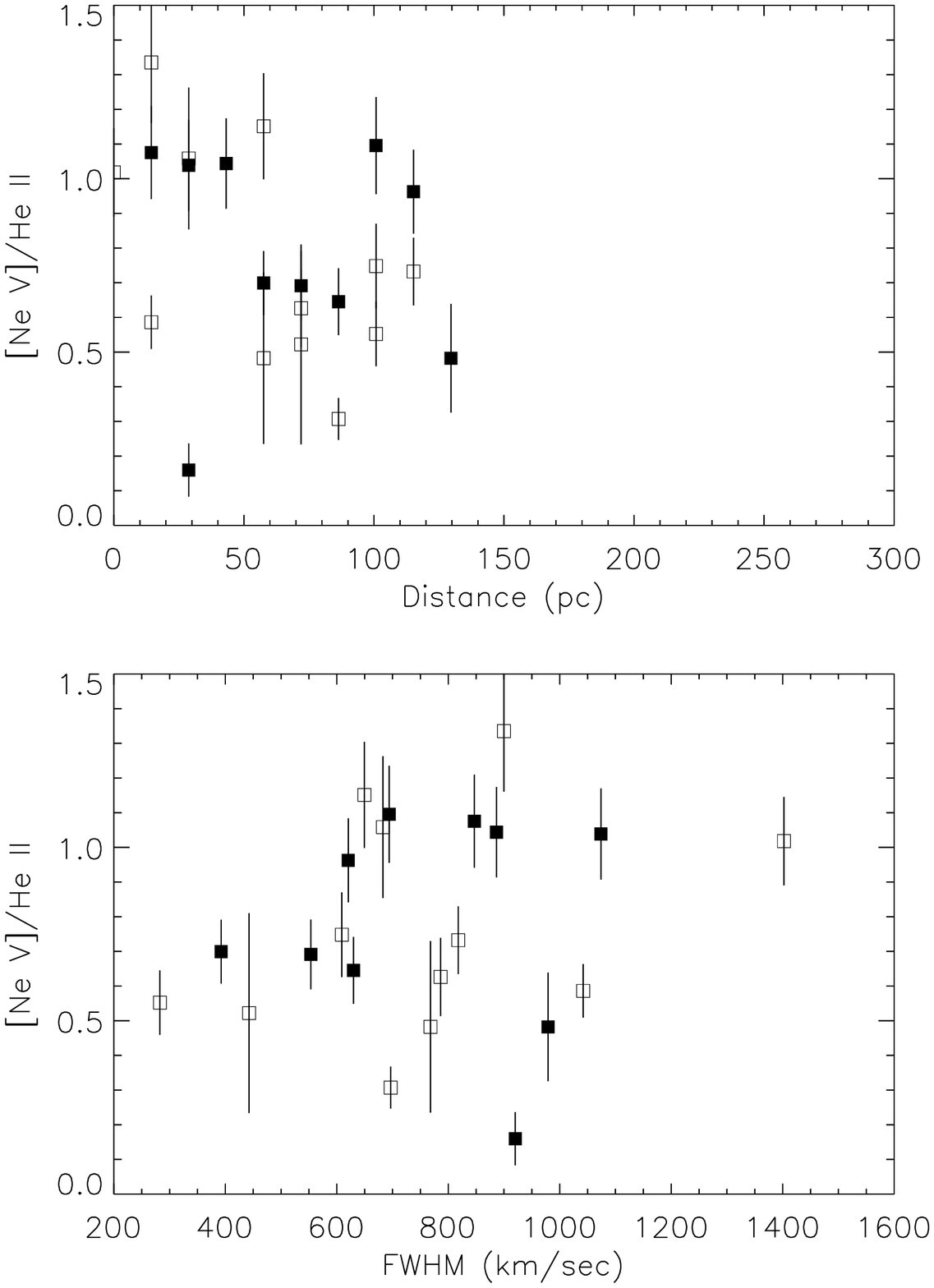}
\\Fig.~3.
\end{figure}

\clearpage
\begin{figure}
\plotone{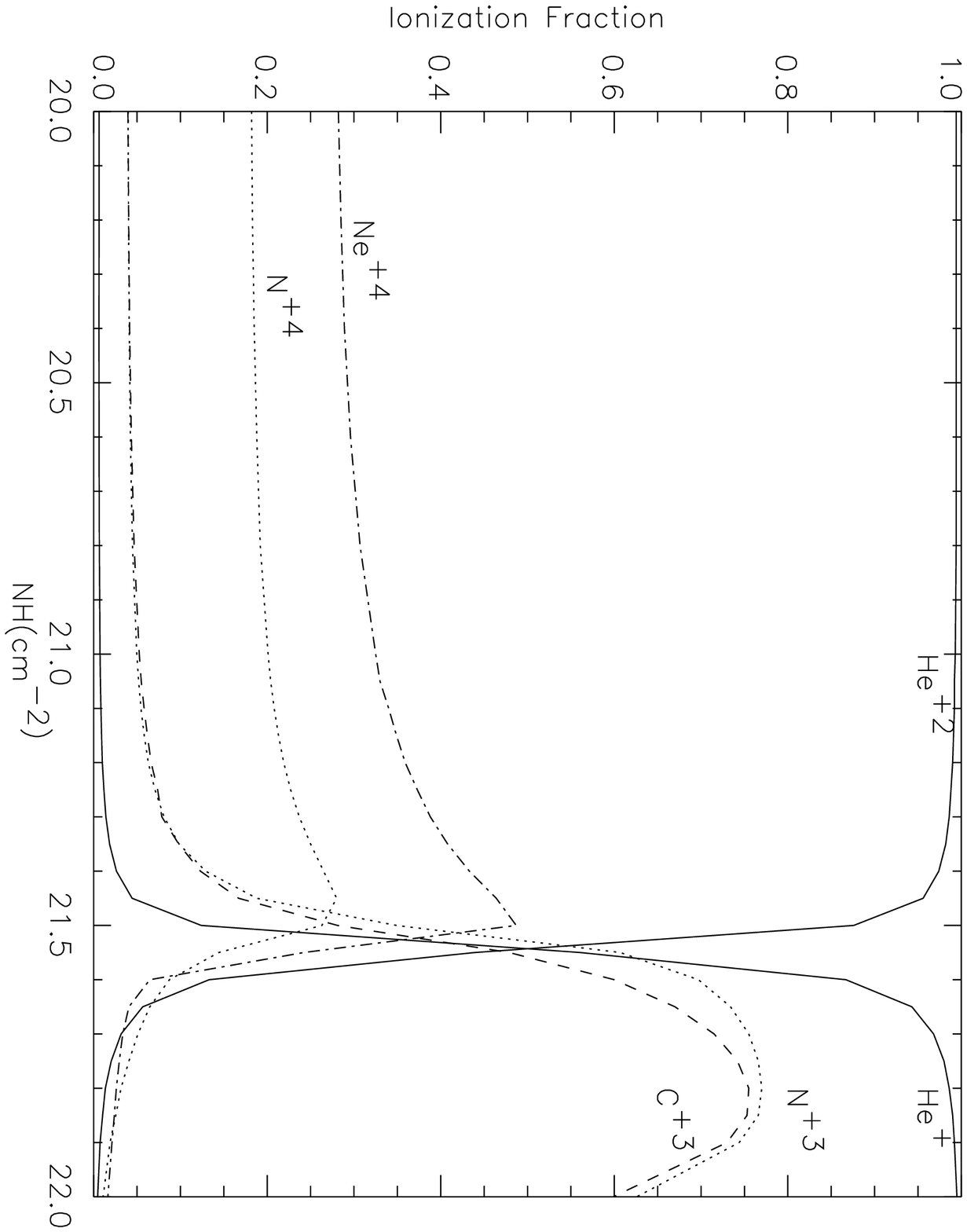}
\\Fig.~4.
\end{figure}

\clearpage
\begin{figure}
\plotone{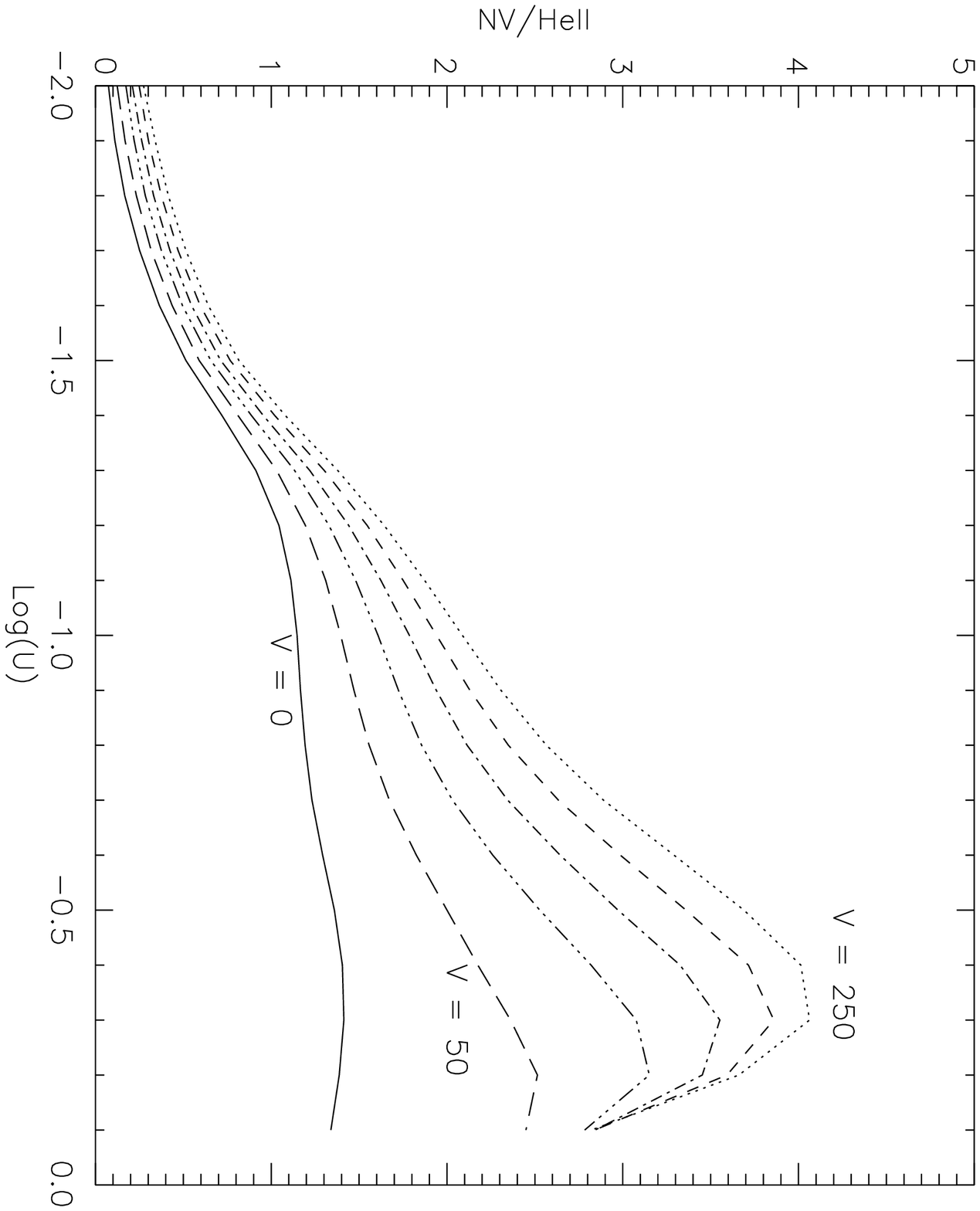}
\\Fig.~5.
\end{figure}

\clearpage
\begin{figure}
\plotone{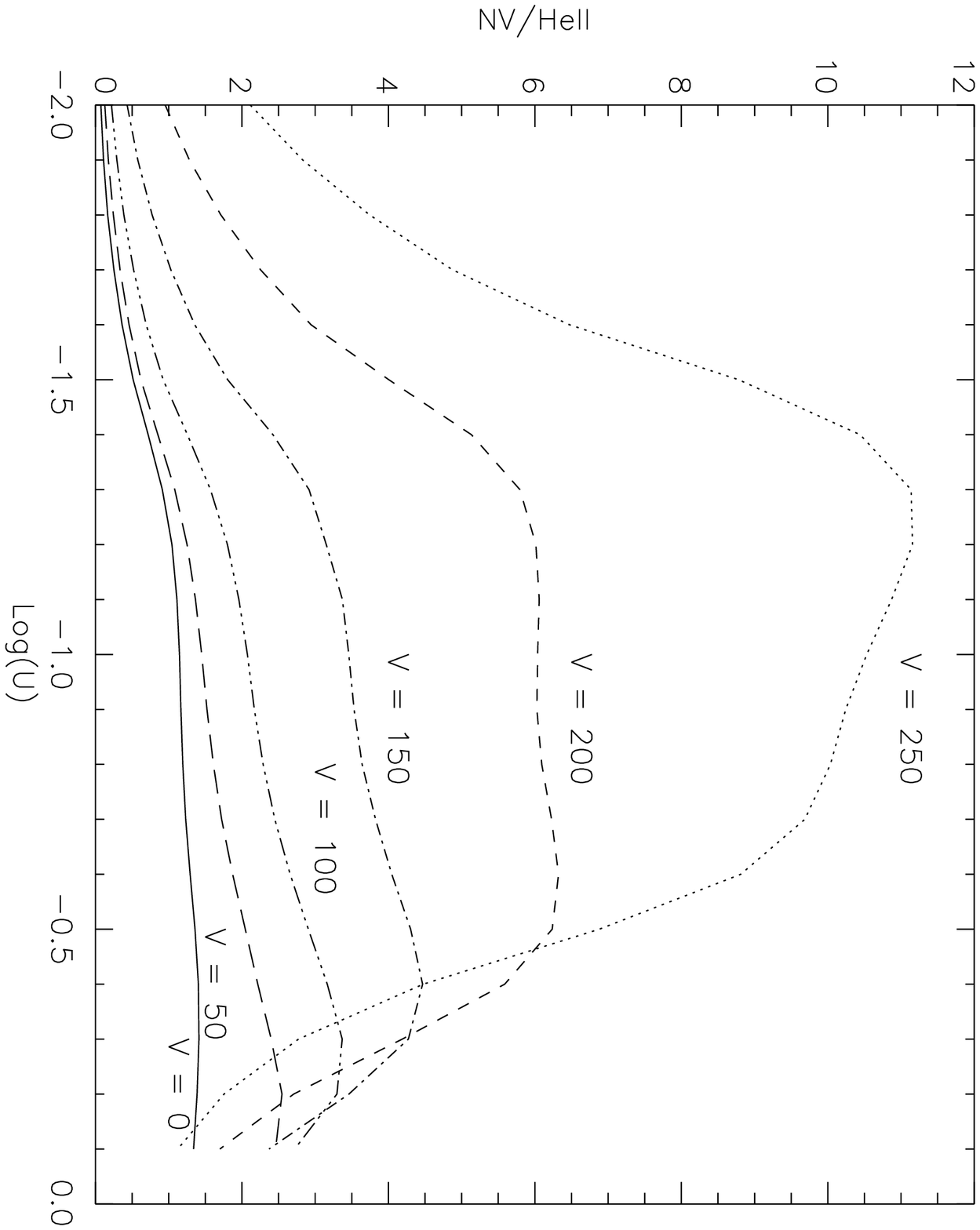}
\\Fig.~6.
\end{figure}

\clearpage
\begin{figure}
\plotone{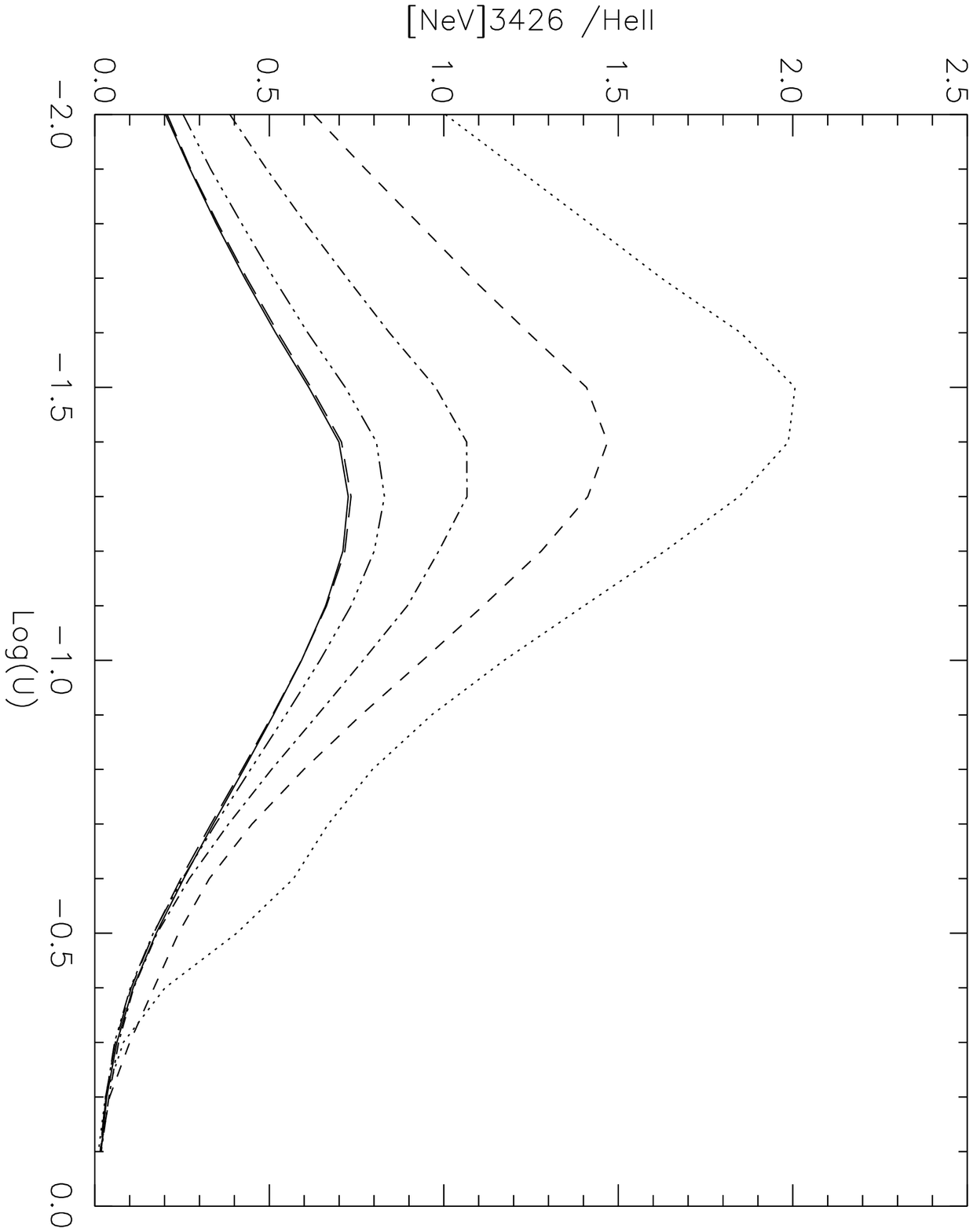}
\\Fig.~7.
\end{figure}

\clearpage
\begin{figure}
\plotone{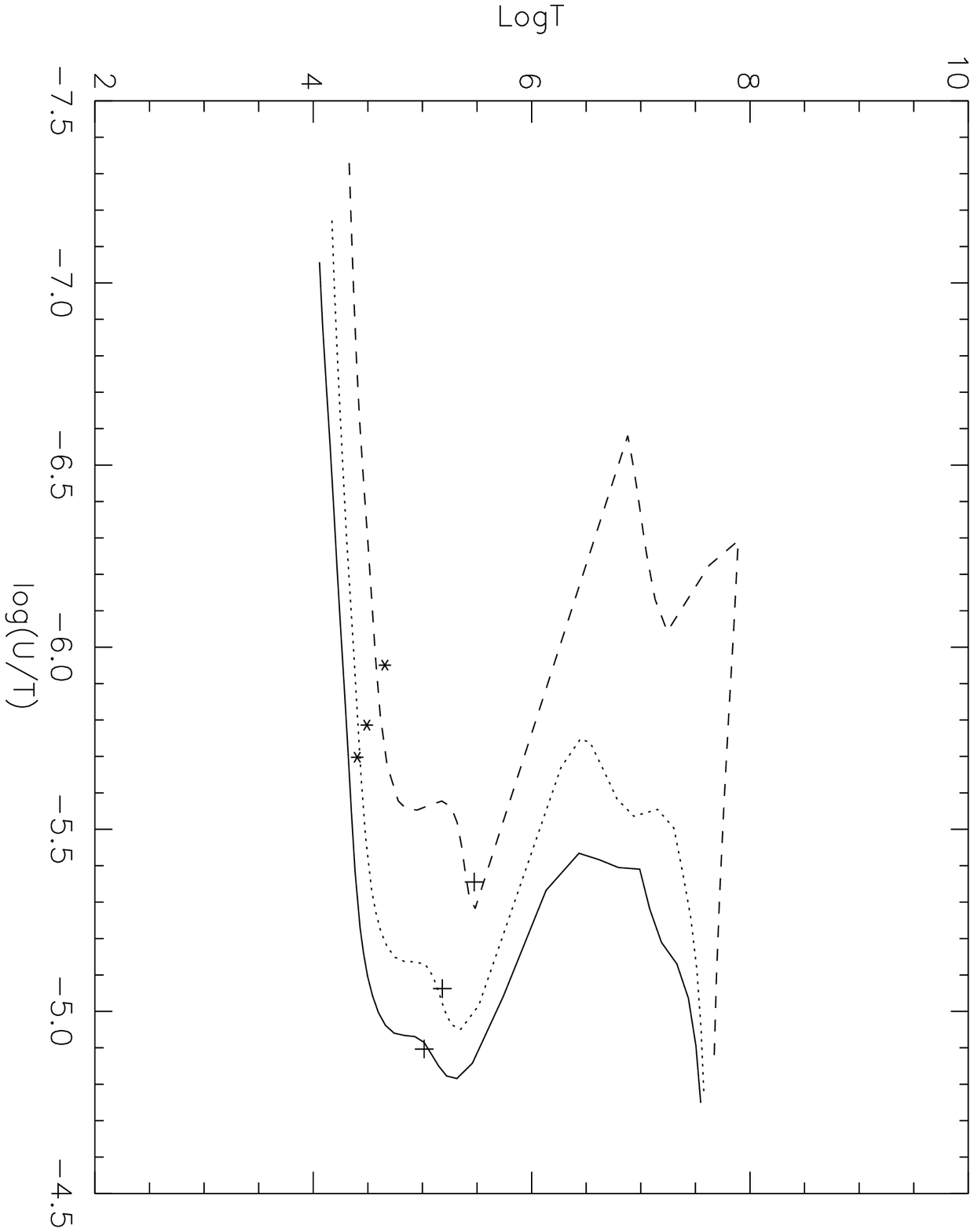}
\\Fig.~8
\end{figure}

\clearpage
\begin{figure}
\plotone{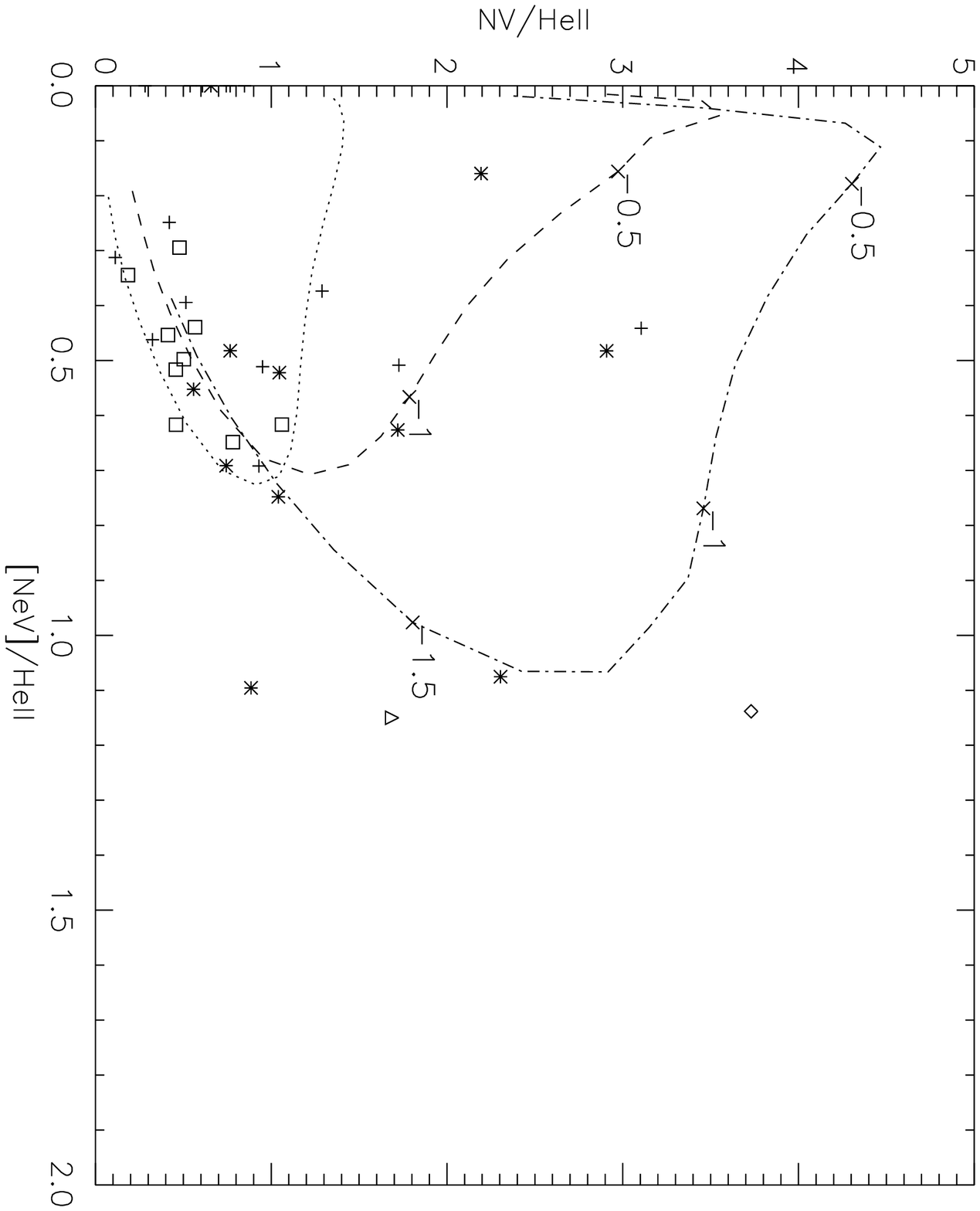}
\\Fig.~9.
\end{figure}

\clearpage
\begin{figure}
\plotone{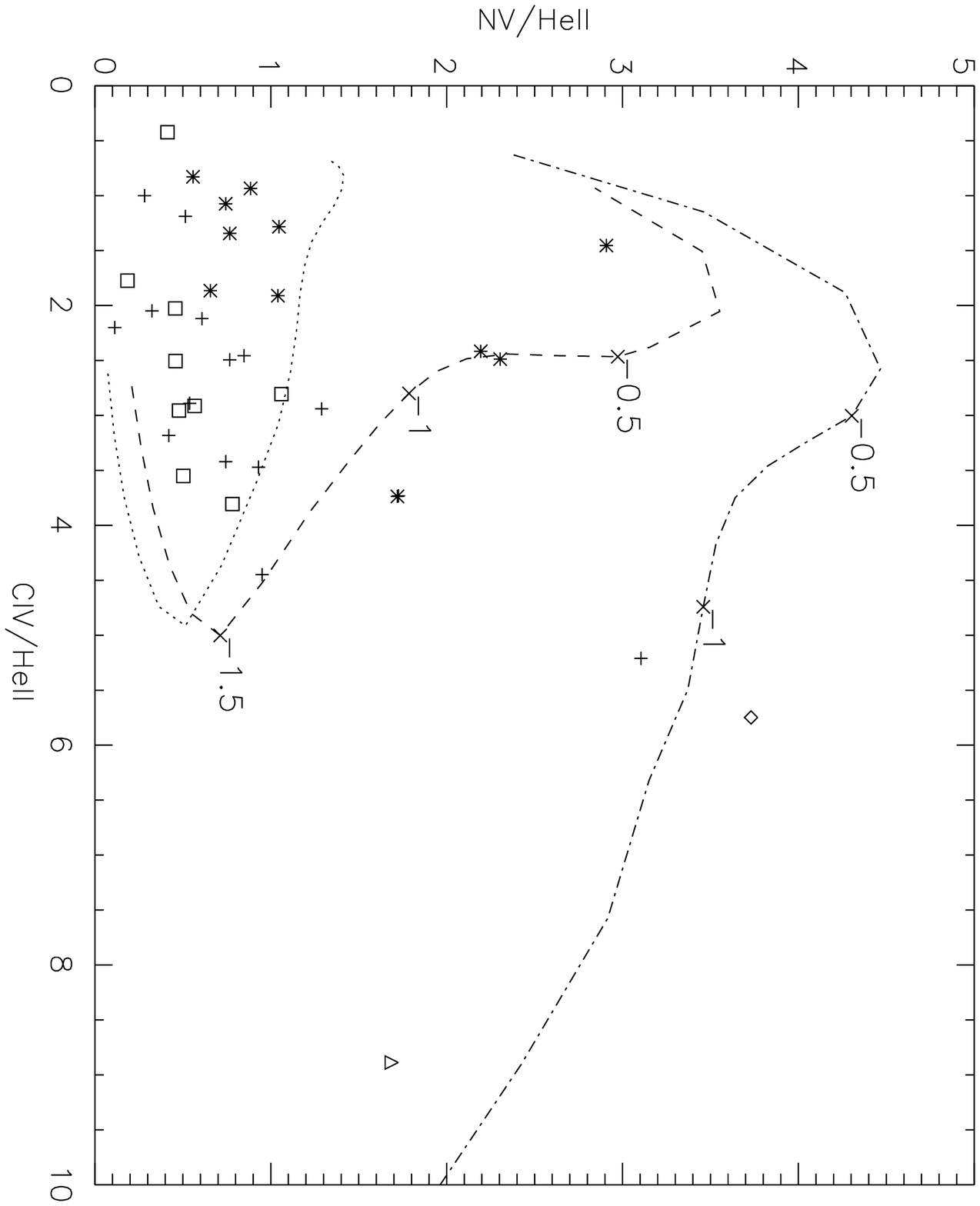}
\\Fig.~10.
\end{figure}


\begin{references}

\reference{all1984}Allen, A.J. 1984, MNRAS, 210, 147

\reference{ara2007}Arav, N., et al. 2007, \apj, 658, 829

\reference{ant1993}Antonucci, R.R.J. 1993, ARA\&A, 31, 473

\reference{arm2007}Armentrout, B.K., Kraemer, S.B., \& Turner, T.J. 2007, \apj,
in press

\reference{asp2005}Asplund, M., Grevesse, N., \& Sauval, A.J. 2005, in ASP
Conference Series 336, Cosmic Abundances as Records of Stellar Evolution and
Nucleosyntheses, ed. T.G. Barnes, III, \& F.N. Bash (San Francisco:ASP), 25

\reference{bla1982}Blandford, R.D. \& Payne, D.G. 1982, MNRAS, 199,883

\reference{bot2000}Bottorff, M.C., \& Ferland, G.J. 2000, MNRAS, 316, 103

\reference{bot2002}Bottorff, M.C., \& Ferland, G.J. 2002, \apj, 568, 581 (BF2002)

\reference{bot2000}Bottorff, M. C., Korista, K.T., \& Sholsman, I. 2000, ApJ, 537, 134

\reference{blu1975}Blumenthal, G.R., \& Mathews, W.G. 1975, \apj, 198, 517

\reference{col2005}Collins, N.R., Kraemer, S.B., Crenshaw, D.M.,
Ruiz, J.R., Deo, R., \& Bruhweiler, F.C. 2005, \apj, 619, 116

\reference{cre1999}Crenshaw, D.M., \& Kraemer, S.B. 1999, \apj, 521, 572

\reference{cre2000}Crenshaw, D.M., \& Kraemer, S.B. 2000, ApJ, 532, L101

\reference{cre2005}Crenshaw, D.M. \& Kraemer, S.B. 2005, \apj, 625, 680

\reference{cre1999}Crenshaw, D.M., Kraemer, S.B., Boggess, A., Maran, S.P., Mushotzky, R.F.,
\& Wu, C.-C. 1999, ApJ, 516, 750

\reference{cre2003}Crenshaw, D.M., Kraemer, S.B., \& George, I.M. 2003, ARA\&A, 41, 117

\reference{cre1996}Crenshaw, D.M., et al. 1996, \apj, 470, 322

\reference{cre2000}Crenshaw, D.M., Kraemer, S.B., Hutchings, J.B., Bradley, L.D.
II, Gull, T.R., Kaiser, M.E., Nelson, C.H., Ruiz, J.R., \& Weistrop D. 2000, AJ,
120, 1731


\reference{cre2000}Crenshaw, D.M., Kraemer, S.B., Hutchings, J.B., 
Danks, A.C., Gull, T.R., Kaiser, M.E., Nelson, C.H., \& Weistrop, D. 2000,
\apj, 545, L27

\reference{das2005}Das, V., Crenshaw, D.M., Hutchings, J.B., Deo, R.P., 
Kraemer, S.B., Gull, T.R., Kaiser, M.E., Nelson, C.H., \& Weistrop, D.
2005, AJ, 130, 945

\reference{das2007}Das, V., Crenshaw, D.M., \& Kraemer, S.B. 2007, \apj, 656,
699

\reference{das2006}Das, V., Crenshaw, D.M., Kraemer, S.B., \& Deo, R. 2006, AJ, 132, 620

\reference{dav1979}Davidson, K., \& Netzer, H. 1979, Rev. Mod. Physics, 51, 715

\reference{del2006}Delahaye, F., \& Pinsonneault, M.H., 2006, \apj, 649, 529

\reference{die1999}Dietrich, M., Wagner, S.J., \& Courvoisier, T.J.L. 1999, in
ASP Conf. Ser. 175, Structure and Kinematics of Quasar Broad Line Regions,
ed. C.M. Gaskell, W.N. Brandt, M. Dietrich, D. Dultzin-Hacyan, \&
M. Eracleous (San Francisco: ASP), 1

\reference{dunn2007}Dunn, J.P., Crenshaw, D.M., Kraemer, S.B., \& Gabel, J.R.
2007, AJ, in press

\reference{eve2007}Everett, J.E., \& Murray, N. 2007, \apj, 656, 93

\reference{fer1998}Ferland, G.J., Korista, K.T., Verner, D.A., Ferguson, J.W., Kingdon, J.B.,
\& Verner, E.M. 1998, PASP, 110, 749

\reference{hut1998}Hutchings, J.B., et al. 1998, \apj, 492, L115


\reference{gre1989}Grevesse, N., \& Anders, E. 1989, in AIP Conf. Proc. 183, Cosmic
Abundances of Matter, ed. C.J. Waddington (New York:AIP), 1

\reference{gre1998}Grevesse, N., \& Sauval, A.J. 1998, Space, Sci. Rev., 85, 161

\reference{gro2004}Groves, B.A., Dopita, M.A., \& Sutherland, R.S. 2004, ApJS, 153, 9

\reference{kha1971}Khachikian, E. Ye., \& Weedman, D.W. 1971, Astrofizika, 7, 389

\reference{kin2002}Kinkhabwala, A., Sako, M., Behar, E., Kahn, S.M., Paerels,
F., Brinkman, A.C>, Kaastra, J.S., Gu, M.F., \& Liedahl, D.A. 2002, \apj, 575,
732

\reference{kos1978}Koski, A.T. 1978, \apj, 223, 56

\reference{kra2000}Kraemer, S.B., \& Crenshaw, D.M. 2000a, \apj, 532, 256

\reference{kra2000}Kraemer, S.B., \& Crenshaw, D.M. 2000b, \apj, 544, 763

\reference{kra1998}Kraemer, S.B., Crenshaw, D.M., Filippenko, A.V., \& Peterson, B.M. 1998a, \apj,
499, 719

\reference{kra2001}Kraemer, S.B., Crenshaw, D.M., \& Gabel, J.R. 2001, \apj, 557, 30

\reference{kra2000}Kraemer, S.B., Crenshaw, D.M., Hutchings, J.B., Gull, T.R., 
Kaiser, M.E., Nelson, C.H., \& Weistrop, D. 2000, \apj, 531, 278

\reference{kra2001}Kraemer, S.B., Crenshaw, D.M., Hutchings, J.B., 
Danks, A.C., Gull, T.R., Kaiser, M.E., Nelson, C.H., \& Weistrop, D. 2001, 
\apj, 551, 671 

\reference{kra1986}Kraemer, S.B., \& Harrington, J.P. 1986, \apj, 307, 478

\reference{kra1999}Kraemer, S.B., Ho, L.C., Crenshaw, D.M., Shields, J.C.,
\& Filippenko, A.V. 1999, \apj, 520, 564


\reference{kra1998}Kraemer, S.B., Ruiz, J.R., \& Crenshaw, D.M. 1998b, \apj, 508, 232

\reference{kra2005}Kraemer, S.B., et al. 2005, \apj, 633, 693

\reference{kro2001}Krolik, J.H., \& Kriss, G.A. 2001, \apj, 561, 684

\reference{kro1981}Krolik, J.H., McKee, C.F., \& Tarter, C.B. 1981, \apj, 249,
422

\reference{kwa1981}Kwan, J., \& Krolik, J.H. 1981, \apj, 250, 478

\reference{mar1978}Marscher, A.P. 1978, \apj, 225, 725

\reference{mat1976}Mathews, W.G. 1976, \apj, 207, 351

\reference{min1997}Minter, A.H., \& Balser, D.S. 1997, \apj, 484, L133

\reference{nel2000}Nelson, C.H., Weistrop, D., Hutchings, J.B., Crenshaw, D.M.,
Gull, T.R., Kaiser, M.E., Kraemer, S.B., \& Lindler, D. 2000, \apj, 531, 275

\reference{oke1968}Oke, J.B., \& Sargent, W.L.W. 1968, \apj, 151, 807

\reference{oli1997}Oliva, E. 1997, in ASP Conf. Ser. 113, Emission Lines
in Active Galaxies: New Methods and Techniques, ed. B.M. Peterson,
F.-Z. Cheng, \& A.S. Wilson (San Francisco: ASP), 288

\reference{ost1989}Osterbrock, D.E., 1989, Astrophysics of Gaseous Nebulae
and Active Galactic Nuclei (Mill Valley: University Science Books)

\reference{pet2004}Peterson, B.M., et al. 2004, \apj, 613, 682

\reference{rui2001}Ruiz, J.R., Crenshaw, D.M., Kraemer, S.B.,
Bower, G.A., Gull, T.R., Hutchings, J.B., Kaiser, M.E., \& Weistrop D. 
2001, AJ, 122, 2961 

\reference{rui2005}Ruiz, J.R., Crenshaw, D.M., Kraemer, S.B.,
Bower, G.A., Gull, T.R., Hutchings, J.B., Kaiser, M.E., \& Weistrop D. 
2005, AJ, 129, 73 

\reference{sak2000}Sako, M., Kahn, S.M., Paerels, F., \& Liedahl, D.A. 2000,
\apj, 543, L115 

\reference{sch1986}Schiano, A.V.R. 1986, \apj, 302, 95

\reference{sch2003}Schmitt, H.R., Donley, J.L., Antonucci, R.R.J., Hutchings, J.B., \& Kimmey, A.L.
2003, ApJS, 148, 327

\reference{sch1996}Schmitt, H.R., \& Kinney, A.L. 1996, \apj, 463, 498

\reference{smi1993}Smith, S.J. 1993, \apj, 411, 570

\reference{sto1998}Stone, J.M., Ostriker, E.C., \& Gammie, C.F. 1998, \apj, 508, L99

\reference{tur2003}Turner, T.J., Kraemer, S.B., Mushotzky, R.F., George, I.M.,
\& Gabel, J.R. 2003, \apj, 594, 128

\reference{vil1993}Vila-Costas, M.B., \& Edmunds, M.G. 1993, MNRAS, 265, 199

\reference{wey1982}Weymann, R.J., Scott, J.S., Schiano, A.V.R., \& Christiansen, W.A.
1982, \apj, 262, 497

\reference{wil1999}Wilson, A.S., \& Raymond, J.C. 1999, \apj, 513, L115

\reference{woo1998}Woodgate, B.E., et al. 1998, PASP, 110, 1183

\end{references}
\end{document}